\documentclass[preprint]{aastex}

\usepackage{natbib}
\usepackage{amsmath}
\usepackage{epsfig}
\usepackage{graphicx}
\usepackage{natbib}
\usepackage{xspace}
\newcommand{\hlf}{\frac{1}{2}}

\newcommand{\df}{\delta F}
\newcommand{\mdf}{\overline{\df}}
\newcommand{\mtx}[1]{\ensuremath{\mathbf{#1}}}

\newcommand{\Kepler}{\textit{Kepler}\xspace} 
\newcommand{\Kmag}{\textit{Kp}\xspace} 

\begin{document}
\shortauthors{E. Petigura \& G. Marcy}

\shorttitle{} 
\title{Identification and Removal of Noise Modes in \Kepler Photometry}

\author{Erik~A.~Petigura}
\affil{Astronomy Department, University of California,
    Berkeley, CA 94720}
\email{epetigura@berkeley.edu}

\author{Geoffrey~W.~Marcy}
\affil{Astronomy Department, University of California,
    Berkeley, CA 94720}


\begin{abstract}
We present the Transiting Exoearth Robust Reduction Algorithm (TERRA) --- a novel framework for identifying and removing instrumental noise in \Kepler photometry.  We identify instrumental noise modes by finding common trends in a large ensemble of light curves drawn from the entire \Kepler field of view.  Strategically, these noise modes can be optimized to reveal transits having a specified range of timescales.  For \Kepler target stars of low photometric noise, TERRA produces ensemble-calibrated photometry having 33~ppm RMS scatter in 12-hour bins, rendering individual transits of earth-size planets around sun-like stars detectable as $\sim 3 \sigma$ signals.
\end{abstract}

\section{Introduction}
The \Kepler Mission is ushering in a new era of exoplanet science.  Landmark discoveries include \Kepler-10b, a rocky planet \citep{Batalha:2011}; the \Kepler-11 system of six transiting planets \citep{Lissauer:2011el}; earth-sized \Kepler-20e and 20f \citep{Fressin12}; KOI-961b, c, and d -- all smaller than earth \citep{Muirhead12}; and \Kepler-16b a circumbinary planet \citep{Doyle:2011ev}.  While \Kepler has revealed exciting individual systems, the mission's legacy will be the first statistical sample of planets extending down to earth size and out to 1 AU.  \Kepler is the first instrument capable of answering ``How common are earths?'' --- A question that dates to antiquity.

Planet candidates are detected by a sophisticated pipeline developed by the \Kepler team Science Operations Center.  In brief, systematic effects in the photometry are suppressed by the Pre-search Data Conditioning (PDC) module, the output of which is fed into the Transiting Planet Search (TPS) module.  For further information, see \cite{Jenkins:2010}.  

The \Kepler mission was designed to study astrophysical phenomena with a wide range of timescales, which include 1-hour transits of hot Jupiters, 10-hour transits of planets at 1 AU, and weeklong spot modulation patterns.  The PDC module is charged with removing instrumental noise while preserving signals with a vast range of timescales.  We review sources of instrumental errors in \S~\ref{sec:InstNoise}, highlighting the effects that are most relevant to transit detection.

The \Kepler team has released candidate planets based on the first 4 and 16 months of data \citep{Borucki11,Batalha12}.  Many of the candidates have additional followup observations from the ground and space aimed at ruling out false positive scenarios.  In addition, statistical arguments suggest that 90-95\% of all candidates and that $\sim 98$\% of candidates in multi-candidate systems are bonafide planets \citep{Morton:2011,Lissauer:2012}.

While \textit{Kepler's} false positive rate is low, its completeness is largely uncharacterized.  If the completeness decreases substantially with smaller planet size or longer orbital periods, the interpretations regarding occurrence drawn from the \cite{Borucki11} and \cite{Batalha12} catalogs will be incorrect.  Hunting for the smallest planets, including earth-sized planets in the habitable zone, will require exquisite suppression of systematic effects.  Without optimal detrending, systematic noise will prevent the detection of the smallest planets, possibly the habitable-zone earth-sized planets, which is the main goal of the \Kepler mission.  Therefore, it is essential for independent groups to develop pipelines that compliment both PDC and TPS.  An early example of an outside group successfully identifying new planet candidates is the Planet Hunters project \citep{Fischer:2011bb,Lintott:2012ut}, which uses citizen scientists to visually inspect light curves.  In addition, existing pipelines from the HAT ground-based search \citep{Huang:2012uj} and the \textit{CoRoT} space mission \citep{Ofir:2012va} have been brought to bear on the \Kepler dataset yielding $\sim100$ new planet candidates.

We present the Transiting Exoearth Robust Reduction Algorithm (TERRA) --- a framework for identifying and removing systematic noise.  We identify systematic noise terms by searching for photometric trends common to a large ensemble of stars.  Our implementation is tuned toward finding trends with transit-length timescales.

\section{Instrumental Noise in \Kepler Photometry}
\label{sec:InstNoise}
\begin{figure}[htbp]
\begin{center}
\includegraphics[width=6.5in]{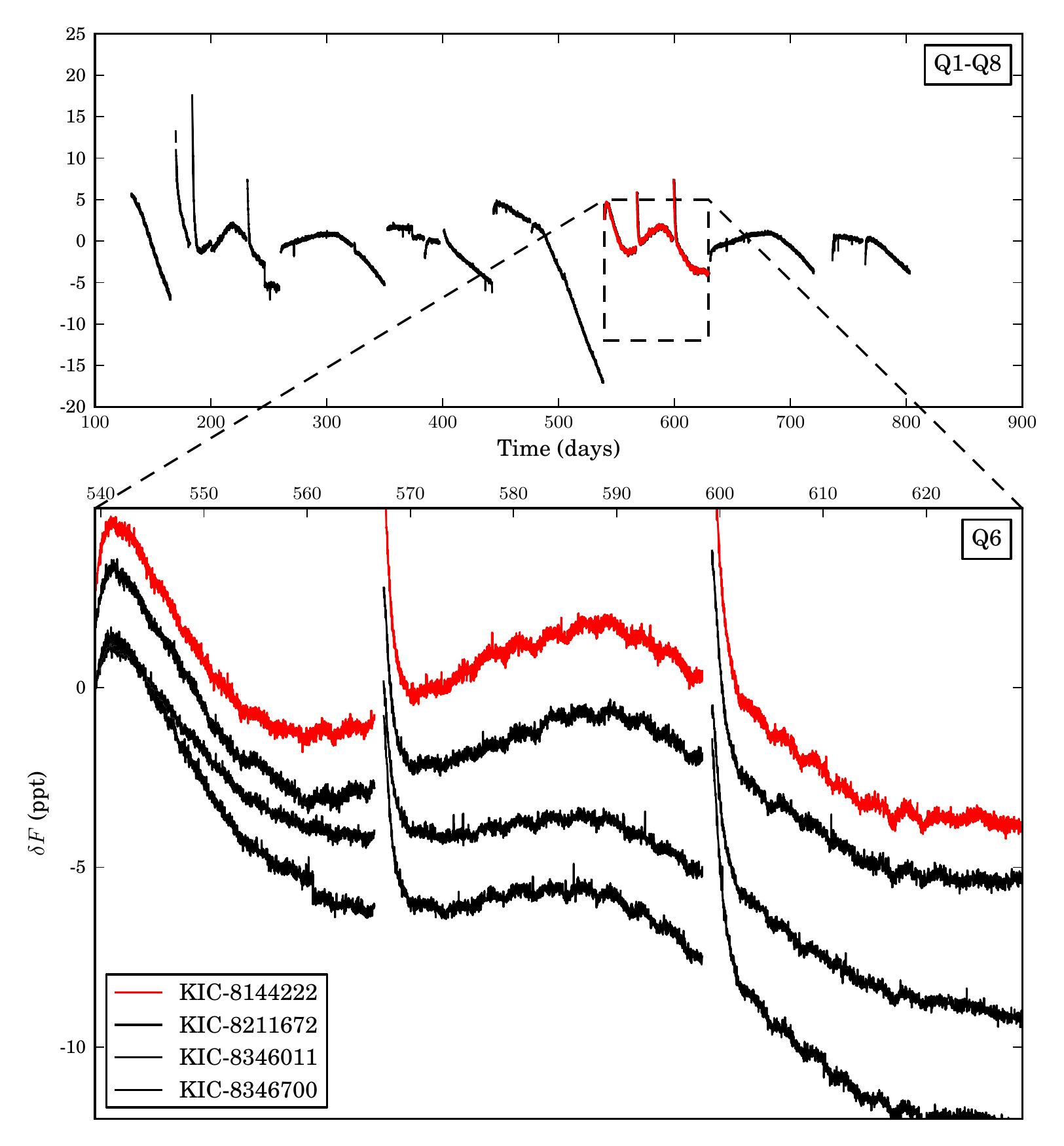}
\caption{Top: Normalized flux from KIC-8144222 (\Kmag=12.4, CDPP12=35.4 ppm) from Quarter 1 through 8 (Q1-Q8).  Bottom: Detail of Q6 photometry showing KIC-8144222 along with three stars of similar brightness, noise level, and location on the FOV (12.0 $<$ \Kmag $<$ 13.0, CDPP12 $<$ 40 ppm, mod.out = 16.1).  Much of the variability is common to the 4 stars and therefore instrumental in origin.  The two spikes are due to thermal settling events, and the three-day ripples are due to onboard momentum management.}
\label{fig:SampLC}
\end{center}
\end{figure}

The \Kepler spacecraft makes photometric observations of $\sim$156,000 targets.  Long cadence photometry is computed by summing all the photoelectrons within a predefined target aperture during a 29.4~minute integration.  The \Kepler team makes this ``Simple Aperture Photometry'' available to the scientific community~\citep{KeplerArchiveManual}.  Simple aperture photometry contains many sources of noise other than Poisson shot noise.  We illustrate several noise sources in Figure~\ref{fig:SampLC}, where we show the normalized photometry ($\df$) of KIC-8144222 (\Kmag = 12.4).  $\df = (F-\overline{F})/\overline{F}$ where $F$ is the simple aperture photometry.

The dominant systematic effect on multi-quarter timescales is ``differential velocity aberration'' \citep{VanCleve:2009}.  As \Kepler orbits the sun, its velocity relative to the \Kepler field changes.  When the spacecraft approaches the \Kepler field, stars on the extremities of the field move toward the center.  Stellar PSFs move over \Kepler apertures by $\sim$ 1 arcsecond resulting in a $\sim$ 1 \% effect over 1-year timescales.

We show a detailed view of KIC-8144222 photometry from Quarter 6 (Q6) in Figure~\ref{fig:SampLC}.  The decaying exponential shapes are caused by thermal settling after data downlinks.  Each month, \Kepler rotates to orient its antenna toward earth.  Since \Kepler is not a uniformly colored sphere, changing the spacecraft orientation with respect to the sun changes its overall temperature.  After data downlink, \Kepler takes several days to return to its equilibrium temperature (Jeffrey Smith, private communication, 2012).  KIC-8144222 photometry also shows a $\sim$0.1\% effect with a 3-day period due to thermal coupling of telescope optics to the reaction wheels.  We explore this 3-day cycle in depth in \S~\ref{sec:interpretation}.

Since all of the previously mentioned noise sources are coherent on timescales longer than one cadence (29.4~minutes), the RMS of binned photometry does not decrease as $1/\sqrt{N}$, where N is the number of measurements per bin.  In order to describe the noise on different timescales, the \Kepler team computes quantities called CDPP3, CDPP6, and CDPP12 which are measures of the photometric scatter in 3, 6, and 12-hour bins.  KIC-8144222 has CDPP12 35.4~ppm and is a low-noise star (bottom 10 percentile).  For a more complete description of noise in \Kepler data see \cite{Christiansen:2011}.

As a comparison, we selected stars which were similar to KIC-8144222 in position on the Field of View (FOV), noise level, and brightness (mod.out = 16.1, CDPP12 $<$ 40 ppm, 12.0 $<$ \Kmag $<$ 13.0).  From this 13-star sample, we randomly selected 3 stars and show their light curves in Figure~\ref{fig:SampLC}.  The photometry from the comparison stars is strikingly similar to the KIC-8144222 photometry.  Since much of the variability is correlated, it must be due to the state of the \Kepler spacecraft.  Common trends among stars can be identified and removed.  The \Kepler team calls this ``cotrending,'' a term we adopt.  

\begin{figure}[htbp]
\begin{center}
\includegraphics[height=6in]{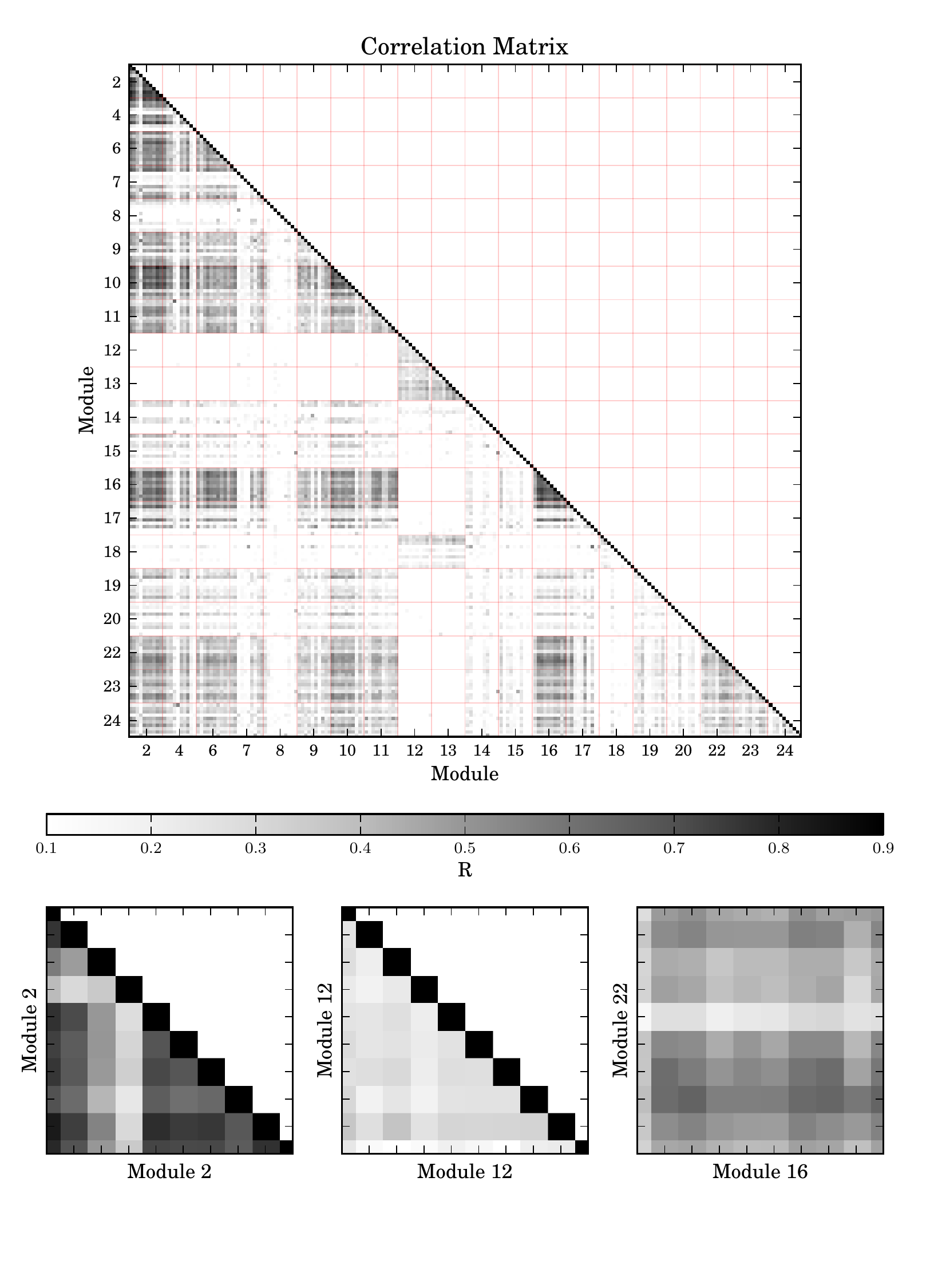}
\end{center}
\caption{Top: Correlation matrix constructed from 200 Q6 light curves.  The correlation (R-value) between two stars is represented by the gray scale, which ranges from 0.1 to 0.9.  The diagonal elements have R = 1.  The stars are ordered according to module and the red lines delineate one module from another.  We enlarge several 10x10 regions in the lower panels.  Stars in some modules (such as module 2) are highly correlated, while other modules (such as module 12) show little correlation.  The module 22 - module 16 correlation matrix is an example of significant intra-module correlation. 
}
\label{fig:corrmat}
\end{figure}

Correlated noise with timescales between 1 and 10 hours can mimic planetary transits and requires careful treatment.  To illustrate the transit-scale correlations among a large sample of stars, we show a correlation matrix constructed from 200 Q6 light curves in Figure~\ref{fig:corrmat}.  The \Kepler photometer is an array of 42 CCDs arranged in 21 modules~\citep{KeplerArchiveManual}.  We organized the rows and columns of the correlation matrix by module.  We constructed the correlation matrix using the following steps:
\begin{enumerate}
\item
We randomly selected 10 light curves from each of the 20 total modules\footnote{Module 3 failed during Q4~\citep{Christiansen:2011}.} from stars with the following properties: $12.5 < $ \Kmag $ < 13.5$ and CDPP12 $< 40$~ppm.  
\item
To highlight transit-scale correlations, we subtracted a best fit spline from the photometry.  The knots of the spline are fixed at 10-day intervals so that we remove trends $\gtrsim$ 10~days.
\item
We normalized each light curve so that its median absolution deviation (MAD) is unity.
\item
We evaluated the pairwise correlation (Pearson-R) between all 200 stars.  
\end{enumerate}
The correlation matrix shows that stars in some modules (e.g. module 2) correlate strongly with other stars in the same module.  However, other modules (e.g. module 12) shows little inter-module correlation.  Finally, the large off-diagonal correlations show that stars in some modules correlate strongly with stars in different modules.  

\section{Identification of Photometric Modes}
We have shown that there is significant high-frequency ($\lesssim$ 10~days) systematic noise in \Kepler photometry.  In order to recover the smallest planets, this noise must be carefully characterized and removed.  We isolate systematic noise by finding common trends in a large ensemble of stars.  This is an extension of differential photometry, widely used by ground-based transit surveys to calibrate out the time-variable effects of the earth's atmosphere.  We find these trends using Principle Component Analysis (PCA).  This is similar to the Sys-Rem, TFA, and PDC algorithms \citep{Tamuz:2005,Kovacs:2005,Twicken:2010}, but our implementation is different.  We briefly review PCA in the context of cotrending a large ensemble of light curves.

\subsection{PCA on Ensemble Photometry}
Consider an ensemble of N light curves each with M photometric measurements.  We can think of the ensemble as a collection of N vectors in an M-dimensional space.  Each light curve $\df$ can be written as a linear combination of M basis vectors that span the space,  
\begin{eqnarray}
\label{eqn:BV}
\df_{1} & = & a_{1,1} V_1 + \hdots  +  a_{1,M} V_M \notag\\
        &\vdots&\\
\df_{N} & = & a_{N,1} V_1 + \hdots  +  a_{N,M} V_M \notag
\end{eqnarray}
where each of the $V_{j}$ basis vectors is the same length as the original photometric time series.  Equation~\ref{eqn:BV} can be written more compactly as
\[
\mtx{D} = \mtx{A} \mtx{V}
\]
where
\[
\mtx{D} =  
\begin{pmatrix}
  \df_{1} \\
  \vdots  \\
  \df_{N} \\
\end{pmatrix},
\mtx{A} = 
\begin{pmatrix}
  a_{1,1} & \hdots & a_{1,M} \\
  \vdots  & \ddots & \vdots  \\
  a_{N,1} & \hdots & a_{N,M} \\
\end{pmatrix},
\mtx{V} = 
\begin{pmatrix}
  V_{1} \\
  \vdots       \\
  V_{M} \\
\end{pmatrix}
\]
Singular Value Decomposition (SVD) simultaneously solves for the basis vectors $\mtx{V}$ and the coefficient matrix $\mtx{A}$ because it decomposes any matrix $\mtx{D}$ into
\[
\mtx{D} = \mtx{USV}^{\mathsf{T}}.
\]
$\mtx{V}$ is an M x M matrix where the columns are the eigenvectors of $\mtx{D}^{\mathsf{T}}\mtx{D}$ or ``principle components,'' and the diagonal elements of S are the corresponding eigenvalues.  The eigenvalues $\{s_{1,1}, \dots, s_{M,M}\}$ describe the extent to which each of the principle components capture variability in the ensemble and are ordered from high to low.  The columns of $\mtx{U}$ are the eigenvectors of  $\mtx{D}\mtx{D}^{\mathsf{T}}$.  Both $\mtx{U}$ and $\mtx{V}$ are unitary matrices, i.e. $\mtx{U}\mtx{U}^{\mathsf{T}} = \mtx{I}$ and $\mtx{V}\mtx{V}^{\mathsf{T}} = \mtx{I}$.

As we saw in \S~\ref{sec:InstNoise}, stars show common photometric trends due to changes in the state of the \Kepler spacecraft.  The most significant principle components will correspond to these common trends.  If we identify the first $N_{Mode}$ principle components as instrumental noise modes, we can remove them via
\begin{equation}
\df_{i,cal} = \df_{i} - \sum_{j=1}^{N_{Mode}} a_{i,j} V_{j}
\label{eqn:fit}
\end{equation}
where $\df_{cal}$ is an ensemble-calibrated light curve.  However since the collection of $\{V_{i}, \hdots V_{M}\}$ spans the space, the higher principle components describe astrophysical variability, shot noise, and exoplanet transits.  We must be careful not to remove too many components because we would be removing the signals of interest.

\subsection{PCA implementation}
\begin{figure}
\begin{center}
\includegraphics{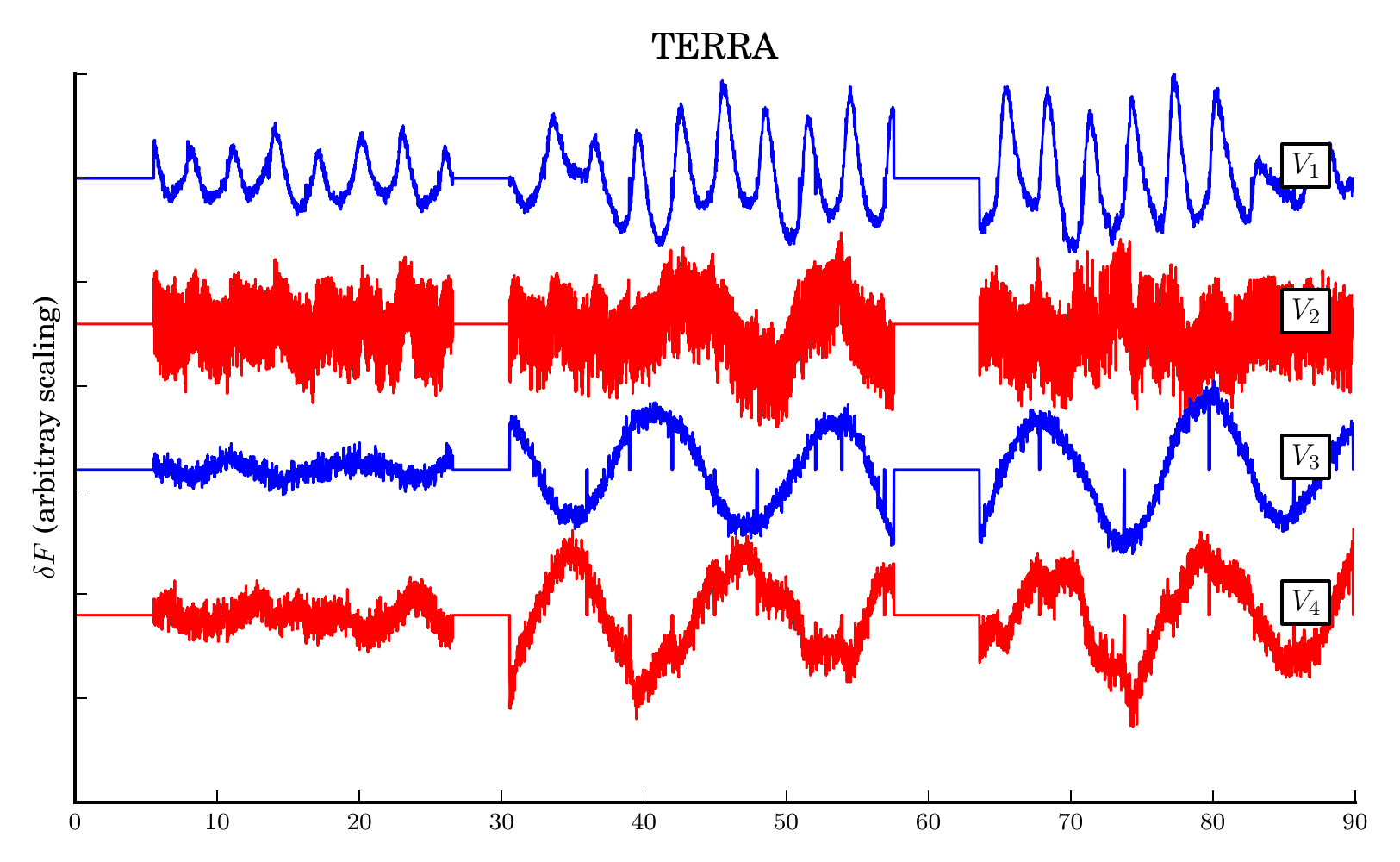}
\end{center}
\caption{
Top: The first four TERRA principle components in our 1000-light curve ensemble plotted in order of significance.  $V_{1}$ has a 3-day periodicity and is due to changes in the thermal state of the spacecraft caused by a 3-day momentum management cycle.  $V_{2}$ has a high frequency component (P = 1.68~hours) that could be due to a 20~minute thermal cycle from an onboard heater aliased with the 29.4~minute observing cadence  alias of a 20-minute thermal cycle driven by an onboard heater.
}
\label{fig:CompCBV}
\end{figure}

We construct a large reference ensemble of light curves $\{\df_{1}, \dots, \df_{N}\}$ of 1000 stars ($12.5 < $ \Kmag $ < 13.5$, CDPP12 $< 40$~ppm) drawn randomly from the entire FOV.  Before performing SVD, we remove thermal settling events and trends $\gtrsim$ 10~days as described in \S~\ref{sec:InstNoise}.  Since SVD finds the eigenvectors of $\mtx{D}^{\mathsf{T}}\mtx{D}$ it is susceptible to outliers as is any least squares estimator.  We perform a robust SVD that relies on iterative outlier rejection following these steps:
\begin{enumerate}
\item
Find principle components and weights for light curve ensemble.
\item
The $i^{\text{th}}$ light curve is considered an outlier if any of the mode weights ($a_{i,1},\hdots,a_{i,4}$) differ significantly from the typical mode weight in the ensemble.  We consider $a_{i,j}$ to be significantly different from the ensemble if  
\[
\frac{|a_{i,j} - \text{med}(a_{j})|}{\text{MAD}(a_{j})} > 10
\]
where med$(a_{j})$ and MAD$(a_{j})$ are the median value and the median absolute deviation of all the $a_{j}$ mode weights.
\item
Remove outlier light curves from the ensemble.
\item
Repeat until no outliers remain.
\end{enumerate}
For our 1000-star sample we identified and removed 51 stars from our ensemble.  These stars tended to have high amplitude intrinsic astrophysical variability, i.e. due to spots and flares.  We plot the four most significant TERRA principle components in Figure~\ref{fig:CompCBV} and offer some physical interpretations of the mechanisms behind these modes in the following section.

\subsection{Interpretation of Photometric Modes}
\label{sec:interpretation}
In this section, we associate the variability captured in the principle components to changes in the state of the \Kepler spacecraft that couple to photometry.  The three-day cycle isolated in our first principle component is due to a well-known, three-day momentum management cycle on the spacecraft \citep{Christiansen:2011}.  To keep a fixed position angle, \Kepler must counteract external torques by spinning up reaction wheels.  These reaction wheels have frictional losses which leak a small amount of heat into the spacecraft, which changes the PSF width and shape of the stars.  

\begin{figure}
\begin{center}
\includegraphics{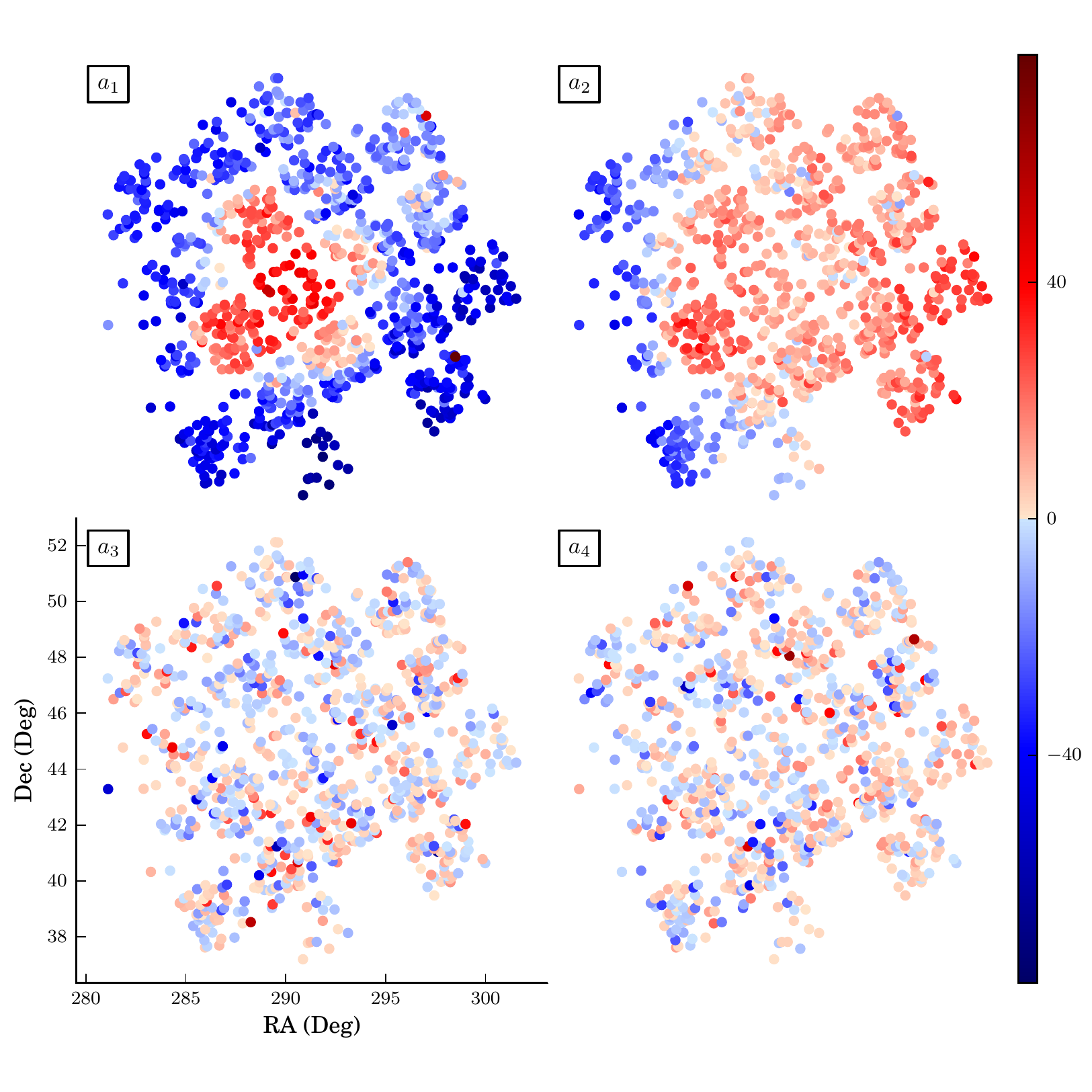}
\end{center}
\caption{
The RA and Dec positions of our 1000-star ensemble.  The points are color-coded by $a_{i}$, the weights for mode $V_{i}$.  Negative values are shown in blue and positive values are shown in red.  The fact that the sign and magnitude of $a_{1}$ depends on distance from the center of the FOV supports the idea that the variability captured by $V_{1}$ is due to PSF breathing of the telescope which is driven by the three-day momentum management cycle.  The gradient in $a_{2}$ could be due to the thermal coupling of an onboard heater to the optics in a tip/tilt sense.  Mode weights $a_{3}$ and $a_{4}$ show no spatial correlation and do not seem to depend on changes in the PSF width.
}
\label{fig:RADecCoeff}
\end{figure}

We can gain a more detailed understanding of this effect, by examining how the mode weights for each reference star corresponding to $V_{1}$, i.e. $\{a_{1,1}, \dots, a_{N,1} \}$, vary across the FOV.  We display the RA and Dec positions of our 1000-star sample in Figure~\ref{fig:RADecCoeff} and color-code the points with the value of $a_{i}$.  The $a_{1}$ and $a_{2}$ mode weights show remarkable spatial correlation across the FOV.  That $a_{1}$ is positive in the center of the FOV and negative at the edges of the FOV means the systematic photometric errors in these two regions respond to the momentum cycle in an anticorrelated sense.  The telescope is focused such that the PSF is sharpest at intermediate distances from the center of the FOV.  Since stars in the center and on the extreme edges have the blurriest PSFs \citep{VanCleve:2009}, they respond most strongly to the momentum cycle. 

The mechanism behind the variability seen in $V_{2}$ is less clear.  $V_{2}$ includes a high frequency component with a period of 1.68 hours.  The \Kepler team has also noticed this periodicity in the pixel scale (Douglas Caldwell, private communication, 2012).  A possible explanation is thermal coupling of the telescope optics to a heater that turns off and on with a $\sim$20~minute period. The 1.68~hour variability would be an alias of this higher frequency with the observing cadence of 29.4~minutes.  The gradient in $a_{2}$ across the FOV suggests the heater is coupled to the telescope optics in a tip/tilt rather than piston sense.  

The higher-order components $a_{3}$ and $a_{4}$ do not show significant spatial correlation, which suggests that $V_{3}$ and $V_{4}$ are not due to changes in the local PSF. Since $V_{3}$ and $V_{4}$ have a $\sim$10-day timescale, they could be the high frequency component of the differential velocity aberration trend that was not removed by our 10-day spline.

\section{Calibrated Photometry}
\subsection{Removal of Modes}
\begin{figure}
\begin{center}
\plotone{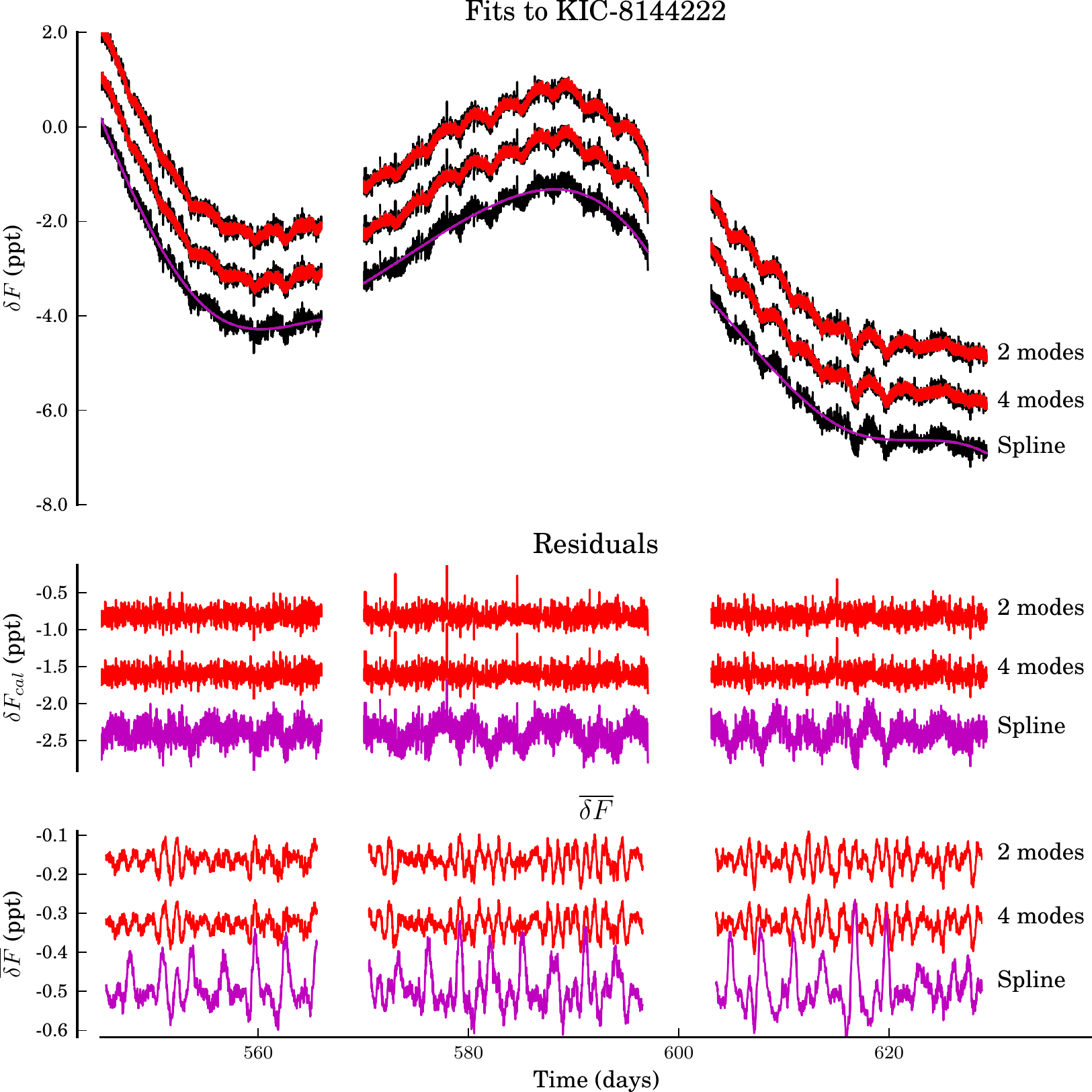}
\end{center}
\caption{
Least squares fits using TERRA principle components to KIC-8144222 Q6 photometry.  The bottom panel shows 12-hour $\mdf$ where smaller scatter implies greater sensitivity to small transits.  We show the spline fit (magenta) as a baseline since it incorporates no ensemble-based cotrending information.  The $\mdf$ using spline detrending shows large spikes at the momentum cycle cusps, which are suppressed in the TERRA cotrending.  Using our robust modes, we are able to produce a clean, calibrated light curve using only two modes.  Decreased model complexity helps guard against overfitting.
}
\label{fig:fits}
\end{figure}

After determining which of the $N_{Mode}$ principle components correspond to noise modes, we can remove them according to Equation~\ref{eqn:fit}.  In Figure~\ref{fig:fits}, we show fits to KIC-8144222 Q6 photometry using different combinations of TERRA principle components.  We achieve uniform residuals using only 2 of our modes as we show quantitatively below.  The simplicity of our model buys some insurance against overfitting. 

\subsection{Performance}
For each of the residuals in Figure~\ref{fig:fits}, we computed the mean depth $\mdf(t_{i})$ of a putative 12-hour transit centered at $t_{i}$ for every cadence in Q6.  The distribution of $\mdf$ due to noise determines the minimum transit depth that can be detected by a transit search algorithm.  $\mdf$ is computed by 
\[
\mdf(t_{i}) = [\df * g](t_{i})
\]
where `$*$' denotes convolution and $g$ is the following kernel
\[
g(t_{i}) = \frac{1}{N_{T}}
	\left[
		\underbrace{\hlf, \dots, \hlf}_{\text{length = }N_{T} },
		\underbrace{-1, \dots, -1}_{\text{length = }N_{T}}, 
		\underbrace{\hlf, \dots, \hlf}_{\text{length = }N_{T} }
	\right].
\]
where $N_{T}$ is 24.  For each of the cotrending schemes, we computed the following statistics describing the distribution of $\mdf$: standard deviation ($\sigma$), 90 percentile (90 \%), and 99 percentile (99 \%).  The standard deviation is roughly equivalent to CDPP12.  Since transit search algorithms key off on peaks in $\mdf$, the percentile statistics are more appropriate figures of merit.  We list these statistics for KIC-8144222 in Table~\ref{tab:fits}.  Ensemble-calibrated photometry produced tighter distributions in $\mdf$ than the spline baseline.

\begin{deluxetable}{l c c c }
\tablewidth{0pc}
\tablecaption{Comparison of fits to KIC-8144222 photometry.}
\tablehead{
\colhead{Cotrending}&
\colhead{$\sigma$} &
\colhead{90 \%} & 
\colhead{99 \%} \\
}
\startdata
2 PMs  & 24  & 28 & 53 \\
4 PMs  & 24  & 28 & 53 \\
Spline & 53  & 66 & 146 \\
\enddata
\tablecomments{Standard deviation, 90 percentile, and 99 percentile (in ppm) of the $\mdf$ distributions for KIC-8144222 using different cotrending schemes.  The spline fit is included as a baseline since it incorporates no ensemble-based cotrending information.  In computing $\mdf$, we have assumed a 12-hour transit duration.  All cotrending approaches yield tighter $\mdf$ distributions than the spline baseline.
}
\label{tab:fits}
\end{deluxetable}

\subsection{Comparison to PDC}
In this section, we offer some simple comparisons between TERRA and the PDC implementation of~\cite{Twicken:2010}.  This paper represents our efforts to improve upon that algorithm.  The \Kepler PDC pipeline has evolved beyond that presented in \cite{Twicken:2010} culminating with PDC-MAP \citep{Stumpe:2012,Smith:2012}.  We feel that the \cite{Twicken:2010} algorithm is an important touchstone for comparison given that the most recent release of planets \citep{Batalha12} was based on photometry that was largely processed with the \cite{Twicken:2010} algorithm.

We assess cotrending performance in the context of transit detectability.  We note that PDC outputs are not directly used in transit detection.  PDC light curves are subject to additional detrending (mostly of low frequency content) before the transiting planet search is run \citep{Tenenbaum:2010}.

\begin{figure}
\begin{center}
\plotone{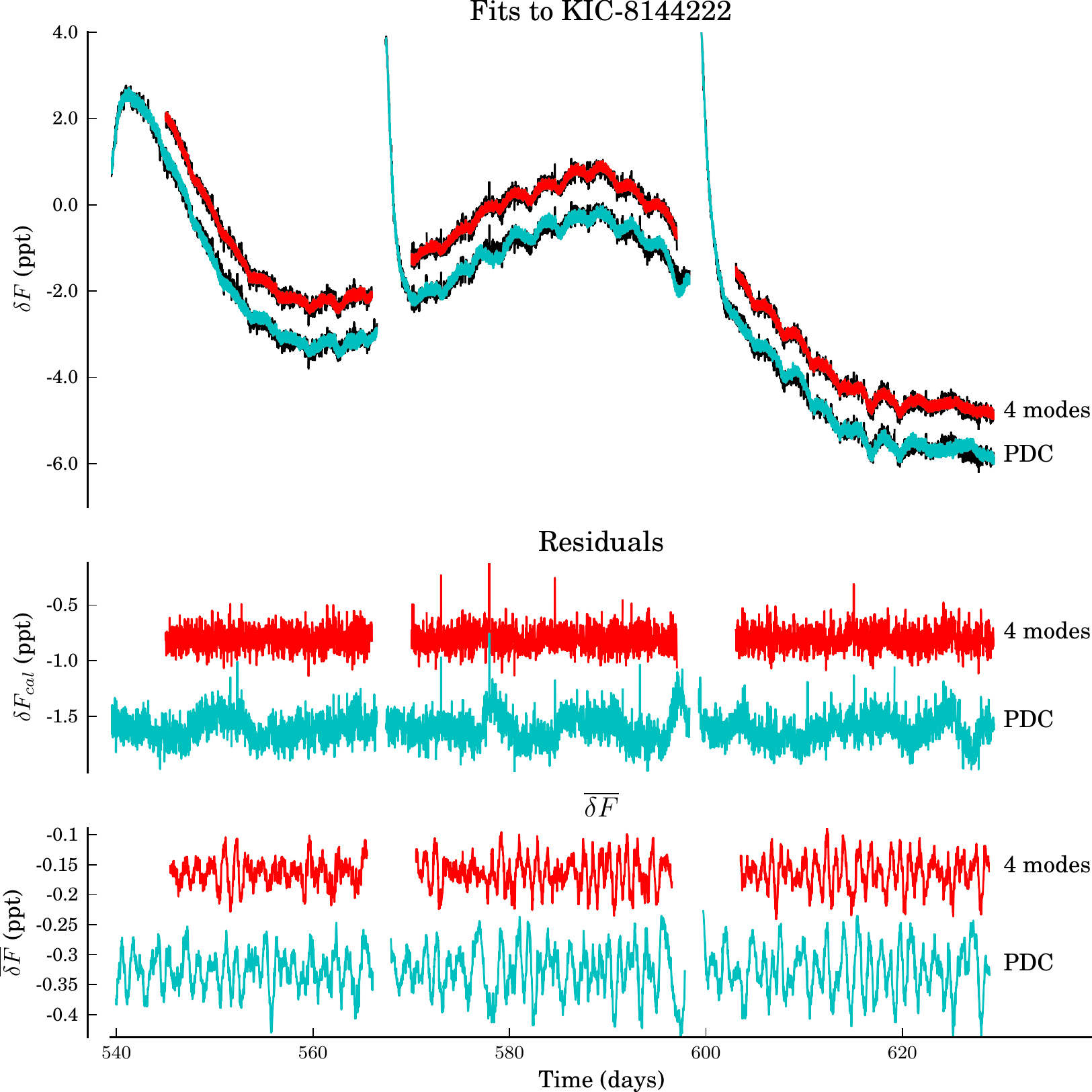}
\end{center}
\caption{
Same as Figure~\ref{fig:fits} except we compare fits using the 4 TERRA modes, with the PDC processed photometry.  The bottom panel shows 12-hour $\mdf$ where smaller scatter implies greater sensitivity to small transits.  The RMS scatter in the 12-hour $\mdf$ distribution is 24~ppm for TERRA processed photometry and 36~ppm for PDC processed photometry.
}
\label{fig:fits2}
\end{figure}

In Figure~\ref{fig:fits2}, we show fits to the KIC-8144222 photometry using 4 TERRA modes and the PDC algorithm.  While PDC flattens photometry collected during the thermal transients, it injects high frequency noise into regions that are featureless in the TERRA-calibrated photometry.  For KIC-8144222, the RMS scatter in the 12-hour $\mdf$ distribution is 24~ppm for TERRA processed photometry and 36~ppm for PDC processed photometry.

\begin{figure}
\begin{center}
\includegraphics{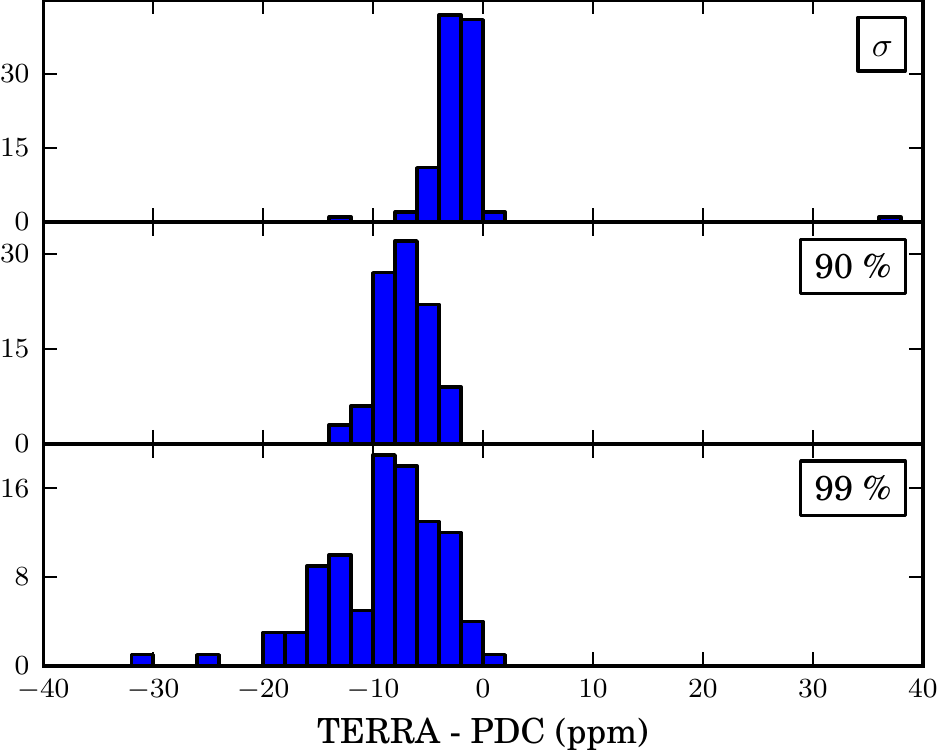}
\end{center}
\caption{
We computed the standard deviation, 90 percentile, and 99 percentile (in ppm) of 12-hour $\mdf$ for 100 light curves using TERRA and PDC cotrending.  The histograms show the difference of the TERRA and PDC statistics.  Negative values mean a tighter $\mdf$ distribution using our cotrending and hence a lower noise floor in a transit search.}
\label{fig:stat100}
\end{figure}

Using 4 TERRA modes, we cotrend 100 stars selected at random from our 1000-star reference ensemble.  We then compute 3, 6, and 12-hour $\mdf$ from TERRA and PDC calibrated light curves.  We then calculate the difference between the $\sigma$, 90\%, and 99\% statistics for TERRA and PDC cotrending.  We show the distribution of these differences for the 12-hour $\mdf$ in Figure~\ref{fig:stat100}.  The median improvement in $\sigma$, 90\%, and 99\% using TERRA cotrending is 2.8, 6.6, and 8.7~ppm.  We tabulate the median values of the $\sigma$, 90\%, and 99\% statistics in Table~\ref{tab:stat100}.

We believe that these comparisons are representative of the stars from which we constructed our reference ensemble ($12.5 < $ \Kmag $ < 13.5$ and CDPP12 $< 40$~ppm).  These bright, low-noise stars are the most amenable to exoearth detection. Our comparisons do not pertain to stars with different brightness or noise level. 

\begin{deluxetable}{l c c c c c c}
\tablewidth{0pc}
\tablecaption{Comparison of TERRA and PDC cotrending performance for 100 stars.}
\tablehead{
\colhead{Transit Width}&
\colhead{$\sigma$}&
\colhead{$\sigma$}&
\colhead{90\%}&
\colhead{90\% }&
\colhead{99\%}&
\colhead{99\% }\\
\colhead{(hours)}&
\colhead{TERRA}&
\colhead{PDC}&
\colhead{TERRA}&
\colhead{PDC}&
\colhead{TERRA}&
\colhead{PDC}\\
}
\startdata
3 & 58 & 60 & 68 & 76 & 129 & 141 \\
6 & 43 & 45 & 50 & 57 & 97 & 105 \\
12 & 33 & 37 & 39 & 47 & 76 & 88 \\
\enddata
\tablecomments{A comparison of the $\mdf$ distributions using TERRA and PDC cotrending of 100 stars drawn randomly from our 1000-star sample.  We have assumed a range of transit widths.  We show the median values of the standard deviation, 90 percentile, and 99 percentile (in ppm) of the $\mdf$ distributions.  For these 100 stars, TERRA yields tighter distributions of $\mdf$.  The improvement ranges from 8 to 12~ppm in the 99 \% statistic.
}
\label{tab:stat100}
\end{deluxetable}

\section{Conclusions}
\label{sec:conclusions}
TERRA is a new technique for using ensemble photometry to self-calibrate instrumental systematics in \Kepler light curves.  We construct a simple noise model by running a high-pass filter and removing thermal settling events before computing principle components.  For a typical $12.5 < $ \Kmag $ < 13.5$ and CDPP12 $< 40$~ppm star, TERRA produces ensemble-calibrated photometry with 33~ppm RMS scatter in 12~hour bins.  With this noise level, a 100~ppm transit from an exoearth will be detected at $\sim 3\sigma$ per transit.

A potential drawback of removing thermal settling events is discarding photometry that contains a transit.  Thermal settling events amounted to 14\% of the valid cadences in Q1-Q8 photometry.  Since signal to noise grows as the square root of the number of transits, removing 14\% of the photometry results in a 7\% reduction in the signal to noise of a given transit.  The completeness of the survey may decrease slightly, since some borderline transits will remain below threshold.  However, this can easily be overcome by gathering 14\% more data.

Ensemble-based cotrending is most effective when the timescales in the ensemble are matched to the signal of interest. We are skeptical that a ``one size fits all'' approach exists and we encourage those who wish to get the most out of \Kepler data to tune their cotrending to the timescale of their signals of interest.

\acknowledgements 
The authors are indebted to Jon M. Jenkins, Andrew W. Howard, Douglas A. Caldwell, Thomas Barclay, Jeffrey C. Smith, and Jeffrey D. Scargle for productive and enlightening conversations that improved this work.  We acknowledge salary support for Petigura by the National Science Foundation through the Graduate Research Fellowship Program.  This work made use of NASA's Astrophysics Data System Bibliographic Services as well as the SciPy \citep{scipy}, IPython \citep{ipython}, and Matplotlib \citep{matplotlib} Python modules.

\bibliographystyle{apj}
\bibliography{manuscript,manuscript_hand,manuscript_bdsk,modes}
\end{document}